\documentclass[lettersize,journal]{IEEEtran}
\usepackage{amsmath,amsfonts}
\usepackage{algorithmic}
\usepackage{algorithm}
\usepackage{array}
\usepackage[caption=false,font=normalsize,labelfont=sf,textfont=sf]{subfig}
\usepackage{textcomp}
\usepackage{stfloats}
\usepackage{url}
\usepackage{verbatim}
\usepackage{graphicx}
\usepackage{cite}
\usepackage{xcolor}
\usepackage{cuted}
\usepackage{makecell}
\usepackage{booktabs}
\usepackage{makecell}
\usepackage[table,xcdraw]{xcolor}

\begin{document}

\title{Computational electro-optic frequency comb spectroscopy}

\author{J.J. Navarro-Alventosa, A. Aupart-Acosta and V. Durán

\thanks{J.J. Navarro-Alventosa, Adrián Aupart-Acosta and Vicente Durán are with the GROC-UJI, Institute of New Imaging Technologies, University Jaume I, 12071 Castellón, Spain}
\thanks{0733-8724 © 2025 IEEE. All rights reserved, including rights for text and data mining, and training of artificial intelligence and similar technologies. Personal use is permitted, but republication/redistribution requires IEEE permission. See https://www.ieee.org/publications/rights/index.html for more information.}}



\maketitle

\begin{abstract}
Computational techniques have gained significant traction in photonics, enabling the co-design of hardware and data processing algorithms to drastically simplify optical system architectures and improve their performance. However, their application in optical frequency comb spectroscopy remains considerably underexplored. In this work, we introduce a non-interferometric approach to frequency comb spectroscopy based on dynamically tailored electro-optic modulation. The core of our method is a reconfigurable electro-optic comb generator capable of producing a sequence of known comb spectra to interrogate a spectroscopic sample. Instead of recording spectrally resolved or interferometric data, our system captures a set of integrated optical power measurements—one per probe comb—from which the sample's spectral response is computationally reconstructed by solving an inverse problem. We present the theoretical foundations of this method, assess its limitations, and validate it through numerical simulations. As a proof of concept, we demonstrate the experimental reconstruction of several spectral signatures, including a molecular absorption line at 1545 nm. For these results, we use numerically computed spectra and experimentally measured power values, all acquired within 10 milliseconds. Finally, we discuss potential extensions and improvements of the method, as well as its integration into chip-scale spectroscopic systems.   
\end{abstract}


\section{Introduction}
\IEEEPARstart{E}{lectro-optic (EO)} modulation of a continuous-wave (cw) laser is a decades-old technique for generating optical frequency combs (OFCs), recently revitalized by its potential for chip-scale integration \cite{bibitem1,bibitem2,bibitem3,bibitem4}. A characteristic feature of EO-OFCs, often overlooked, is that they can be analytically described and accurately simulated \cite{bibitem2}. In a very basic configuration, a comb spectrum can be created using a phase modulator driven by a single-tone radiofrequency (RF) signal. In this case, Bessel functions of increasing order define the amplitude of consecutive spectral lines. Due to the oscillatory properties of these functions, the generated spectrum is not uniform. Indeed, a line can be nearly suppressed if, for a given modulation index, the corresponding Bessel function is close to a null value. Moreover, any change in the driving signal's amplitude redistributes the optical power among the spectral lines, while the total power remains, in principle, constant. Thus, a wide variety of spectra can be produced in a predictable and controllable manner. This ability, however, has usually been considered a disadvantage, and many methods have been proposed to achieve flat-topped EO-OFCs, such as cascaded EO modulators \cite{bibitem5,bibitem6}, dual-drive Mach-Zehnder modulators \cite{bibitem7} and optimized driving RF signals \cite{bibitem8,bibitem9}.

In contrast to other comb platforms, EO-OFCs offer easy implementation, robustness, and a flexibility that enables line spacing tuning across several orders of magnitude (from megahertz to tens of gigahertz). This comb category is also well suited for dual-comb spectroscopy (DCS) \cite{bibitem10}, as the generation of two EO-OFCs from a single cw laser inherently ensures high mutual coherence \cite{bibitem11,bibitem12}. A limitation of EO combs is their modest number of spectral lines, typically less than one hundred. The comb bandwidth, however, can be dramatically increased by non-linear broadening techniques \cite{bibitem11,bibitem13,bibitem14} or by placing the EO modulator within a cavity \cite{bibitem15,bibitem16}, albeit at the expense of reduced flexibility. EO combs have been successfully employed in a wide range of applications, including molecular spectroscopy \cite{bibitem11,bibitem17}, distance measurement \cite{bibitem18,bibitem19}, telecommunications \cite{bibitem20}, and, more recently, distributed fiber sensing \cite{bibitem21,bibitem22}. Beyond these established applications, the particular characteristics of EO combs make them especially appropriate for new computational approaches.

In the field of optical frequency combs, the term 'computational' has primarily been applied to DCS schemes that enable coherent averaging without the need for stabilization \cite{bibitem23,bibitem24}. Other techniques that can be considered computational—such as scanning apodization or compressive sensing—have also been demonstrated for DCS using programmable, self-referenced OFCs, which offer dynamic control of the timing and phase of optical pulses at the attosecond level \cite{bibitem25,bibitem26}. Nevertheless, the complexity of these programmable combs limits their applicability to advanced research laboratories. For this reason, ongoing research is turning to EO combs as a means to simplify such architectures \cite{bibitem27}. In this paper, we present a completely different computational approach that resembles the spectral analogue of a 'single-pixel' camera, an imaging scheme in which an object is illuminated with a series of spatially resolved patterns, and its image is subsequently reconstructed from powers measured by a non-pixelated detector \cite{bibitem28,bibitem29}. In a similar way, we dynamically reconfigure an EO modulator to generate a programmed sequence of comb spectra that repeatedly interrogate a spectroscopic sample (Fig. 1). This process leverages the ability of an EO comb generator to produce dissimilar spectra (in both shape and number of usable lines) by varying the applied RF signal. Each time the comb generator is updated, the transmitted (or reflected) power depends on the degree of similarity between the sample's spectral signature and the probe comb. The signature is then computationally retrieved by solving a linear system that involves the probe combs and the measured powers. 

Our computational approach is, therefore, a readily implementable, non-interferometric, scan-less scheme that enables spectroscopic intensity measurements. It does not rely on a dispersive element to spatially separate and resolve the spectral comb lines \cite{bibitem30}. Unlike dual-comb schemes that employ a single EO modulator \cite{bibitem31,bibitem32}, the total optical bandwidth does not need to be partially filtered, nor is it eventually limited by the modulator bandwidth. Importantly, the fact that EO-OFCs can be accurately simulated allows the generated spectra to be numerically evaluated rather than measured,  provided that a careful electrical characterization of the comb generator is performed. This is especially convenient when dealing with combs with small line spacing (e.g. $<1$ GHz), whose measurement in the optical domain requires an ultra-high-resolution spectrum analyzer or interferometric approaches. From an experimental point of view, and once the generated comb spectra are determined by any given method, the proposed scheme simply implies acquiring a series of optical powers each time the EO modulation is reconfigured (in our experiments, on the scale of $100$ $\mu$s).

Like many EO comb systems with spectral coverage limitations, our approach is well suited to detect and monitor specific spectral signatures \cite{bibitem33}. A priori knowledge of the sample response characteristics, commonly encountered in sensing applications, allows us to adjust both the bandwidth and spectral resolution of the EO combs, as well as to tailor the RF signal design, aiding in the convergence of reconstruction algorithms. As will be shown, if the spectral response exhibits a well-defined symmetry, the RF signals sent to the comb generator are reduced to simple modulation schemes, easily programmable with a conventional RF generator. When a high-performance arbitrary waveform generator is available, the design of the RF signals can be tailored to address more general and complex spectroscopic challenges.

This paper is organized as follows. In Sec. II, we detail the general operation principle of our approach and show its capabilities from numerical simulations. Sec. III presents a proof of concept with experimental results for two types of samples: a narrow spectral bandpass filter and molecular transition of hydrogen cyanide. Finally, in Sec. IV, we provide a brief outlook and discuss potential improvements and future developments.

\begin{figure*}
\centering
\includegraphics[height=8.4cm]{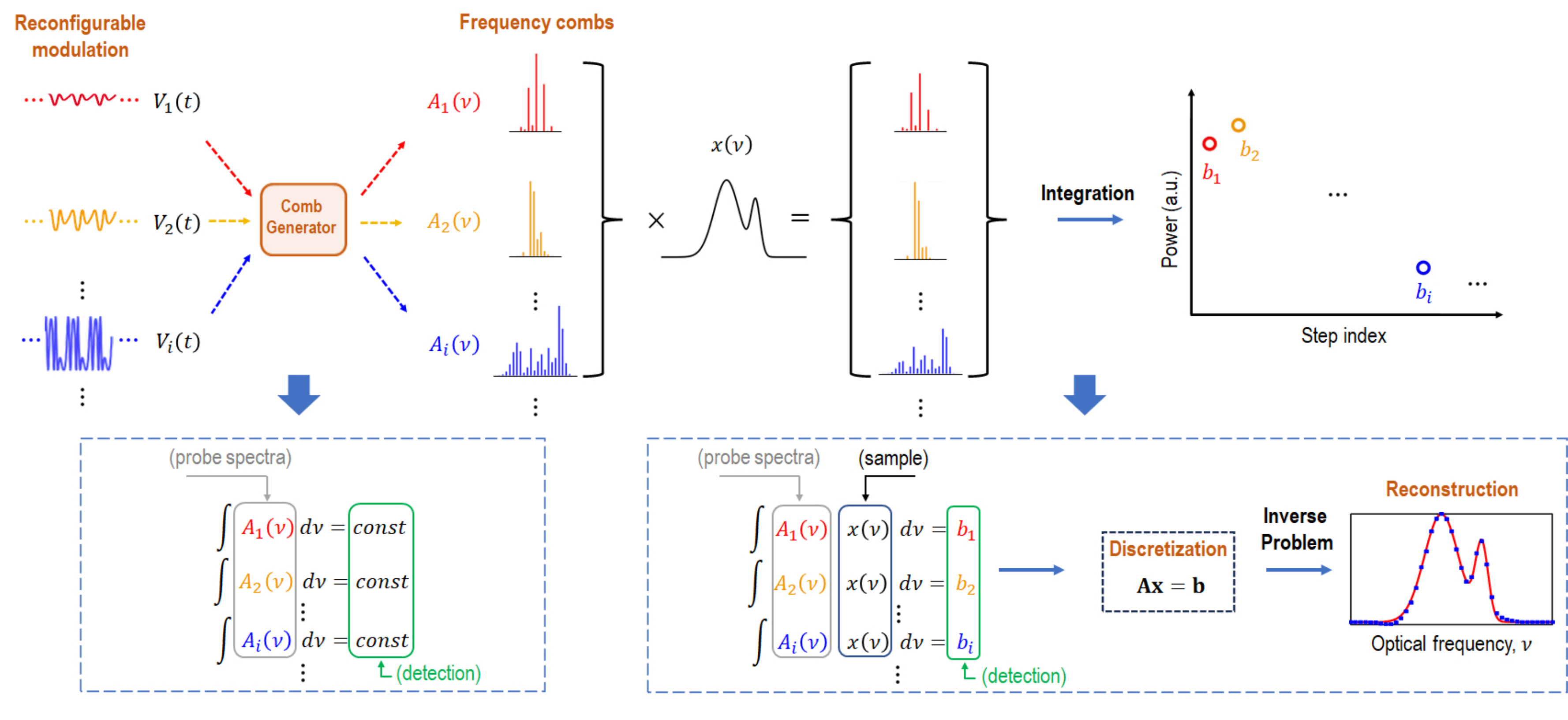}
\caption{\label{fig:2_1_wide}Operational principle. A comb generator, driven by a dynamically reconfigurable signal and fed by a cw laser, generates a set of known comb spectra for interrogating a spectroscopic sample. In each reconfiguration, the power is redistributed among the comb lines but remains constant, as illustrated in the bottom-left inset (\textit{const} denotes a constant value). In contrast, the power transmitted through (or reflected by) the sample varies sequentially based on the overlap between its spectral response and each generated comb. The sample's response is then reconstructed from the measured power values by solving a linear system $\textbf{Ax}=\textbf{b}$, where $\textbf{A}$ is the spectrum matrix, and $\textbf{x}$ and $\textbf{b}$ represent the spectral response and the measured powers, respectively (see bottom-right inset). In practice, this linear system is ill-posed, and its solution requires formulating and solving an optimization problem.}
\end{figure*}

\section{Operation principle}
\subsection{Driving signals}

The key component of our technique is a dynamically reconfigurable EO generator, as shown in Fig. 1. In its simplest configuration, the generator consists of a phase modulator (PM) that produces a sequence of $M$ dissimilar comb power spectra with a line spacing $f_{s}$, $\{A_{i}\left( \nu \right)\}\left( i=1,...,M \right)$, where $\nu$ denotes optical frequency. The driving voltage changes in successive steps, each lasting $\Delta t \gg 1/f_{s}$. For the \textit{i}-th step, we consider a general multitone RF voltage given by

\begin{equation}
V_{i} \left( t \right) = \sum^{h}_{l=1}u_{li}sin\left( 2\pi l f_{s}t + \varphi_{li} \right), \\
\label{Eq_2_1}
\end{equation}

\noindent where $h$ is the total number of harmonics that make up the signal. In the above equation, $u_{li}$ and $\varphi_{li}$ represent the peak voltage and phase of the \textit{l}-th harmonic, respectively. The PM is fed by a continuous-wave (cw) laser with an amplitude $E_{0}$ and a carrier frequency $\nu_{0}$. The EO modulation yields an output optical field $E_{i}$ that can be written as a function of time as 

\begin{equation}
E_{i} \left( t \right) = E_{0}e^{j2\pi \nu_{0}t}e^{j\Phi_{i}\left( t \right)}. \\
\label{Eq_2_2}
\end{equation}

In this expression, $j$ denotes the imaginary unit and $\Phi_{i}\left( t \right)$ is the instantaneous phase acquired by the optical field, which is given by

\begin{equation}
\Phi_{i}\left( t \right) = \sum^{h}_{l=1}\beta_{li}sin\left( 2\pi l f_{s}t + \varphi_{li} \right), \\
\label{Eq_2_3}
\end{equation}

\noindent where $\beta_{li}$ is the modulation index, defined in terms of the half-wave voltage of the PM $V_{\pi}$ as $\beta_{li} \equiv \pi u_{li}/V_{\pi}$ \cite{bibitem1}. Although $V_{\pi}$ remains approximately constant in our experiments, it actually depends on the applied RF frequency and can vary significantly with the harmonic order $l$.

For the results presented in this paper, it suffices to consider up to two harmonics. The more complex case of \textit{h}-tone modulation is treated without any simplifying assumptions in Appendix A.

\subsubsection{\textit{Single-tone RF signal}}
Let us start with a voltage that is a pure sinus $(h=1)$. In this case, the field emerging from the PM can be expressed as a Fourier series expansion, with harmonic components weighted by Bessel functions of different order. Specifically, by applying the Jacobi-Anger identity to the factor $e^{j\Phi_{i}\left( t \right)}$ of Eq. (2), we obtain \cite{bibitem2}

\begin{equation}
E_{i} \left( t \right) = E_{0}e^{j2\pi \nu_{0}t}\left( \sum^{\infty}_{k=-\infty} J_{k}(\beta_{i})e^{j2\pi kf_{s}t} \right), \\
\label{Eq_2_4}
\end{equation}

\noindent where $\left\{ J_{k} \right\}$ are Bessel functions of the first kind. For simplicity, we have omitted the subscript $l$ in the expression for $\Phi_{i}\left( t \right)$ given by Eq. (3) and, without loss of generality, assumed that $\varphi_{i}=0$ $\forall$ $i$. Taking into account $J_{-k}=\left( -1 \right)^{k}J_{k}\left( k>0 \right)$, the output field can be written as:

\begin{equation}
\begin{split}
\label{Eq_2_5}
E_{i} (t) &= E_{0}e^{j2\pi \nu_{0}t} \bigl(
J_{0} (\beta_{i}) + \sum^{\infty}_{k=1} J_{k}(\beta_{i})e^{j2\pi kf_{s}t} \\
&\quad + \sum^{\infty}_{k=1} ( -1 )^{k} J_{k}(\beta_{i})e^{-j2\pi kf_{s}t}
\bigr).
\end{split}
\end{equation}

The Fourier transform of this equation reveals a well-known result: the output power spectrum consists of lines evenly spaced by $f_{s}$ and symmetric about $\nu_{0}$. The comb's shape and total number of lines $N$ (typically defined relative to a minimum power threshold) depend on the modulation index $\beta_{i}$, which can be adjusted by varying the applied peak voltage $u_{i}$. As a result, a sequence of comb spectra can be generated by driving a PM with a sinusoidal signal of frequency $f_{s}$ and an amplitude modulated in a staircase pattern, as we will show in Sec. IIB. In practice, the maximum $\beta_{i}$ is limited by the RF power handling capability of the PM. Figure 2(a) illustrates an example of a comb generated by applying a single-tone signal with a modulation index of $2\pi$.

\subsubsection{\textit{Two-tone RF signal}}
Consider now a voltage composed of two harmonics $(h=2)$. According to Eq. (3), the phase $\Phi_{i}\left( t \right)$ for the \textit{i}-th step is given by

\begin{equation}
\Phi_{i}\left( t \right) = \beta_{1i}sin\left( 2\pi f_{s}t \right) + \beta_{2i}sin\left( 4\pi f_{s}t + \varphi_{2i} \right), \\
\label{Eq_2_6}
\end{equation}

\noindent where we assume that $\varphi_{1i}=0$, allowing $\varphi_{2i}$ to be interpreted as the relative phase between the two tones comprising the voltage signal. For the above two-harmonic phase, it remains possible to express $e^{j\Phi_{i}\left( t \right)}$ in Eq. (2) as a Jacobi-Anger expansion, but using two-variable, one-parameter generalized Bessel functions (GBFs) \cite{bibitem34} as follows:

\begin{equation}
e^{j\Phi_{i}\left( t \right)} = \sum^{\infty}_{k=-\infty}e^{j2\pi kf_{s}t} J_{k}(\beta_{1i},\beta_{2i};e^{j\varphi_{2i}}), \\
\label{Eq_2_7}
\end{equation}

\noindent where $\left\{ J_{k}(\beta_{1i},\beta_{2i};\exp^{j\varphi_{2i}}) \right\}$ are the GBFs, whose definition in terms of ordinary one-variable Bessel functions is given by (see Appendix A):

\begin{equation}
J_{k}(\beta_{1i},\beta_{2i};e^{j\varphi_{2i}}) = \sum^{\infty}_{l=-\infty}e^{jl\varphi_{2i}} J_{k-2l}(\beta_{1i})J_{l}(\beta_{2i}), \\
\label{Eq_2_8}
\end{equation}

In practice, Eq. (8) includes only a finite number of terms, depending on the maximum values that $\beta_{1}$ and $\beta_{2}$ take across a complete series of measurements. The relevant fact here is that, due to the symmetry properties of GBFs, introducing the harmonic at $2f_{s}$ makes the generated comb spectra asymmetric with respect to the carrier frequency $\nu_{0}$. Figures 2(b) and 2(c) illustrate how the comb spectrum in Fig. 2(a) changes when the second harmonic is added with $\beta_{2}=\beta_{1}/10$ and relative phases of $0$ and $\pi$, respectively. A sequence of voltage signals with varying $\beta_{1i},\beta_{2i}$ and $\varphi_{2i}$, consequently, can be employed to generate a variety of dissimilar and asymmetric combs, as observed in the spectra shown in Fig. 1. Additional examples are provided in  Sec. S3 of the Supplementary Material.   

\begin{figure*}
\centering
\includegraphics[height=4.3cm]{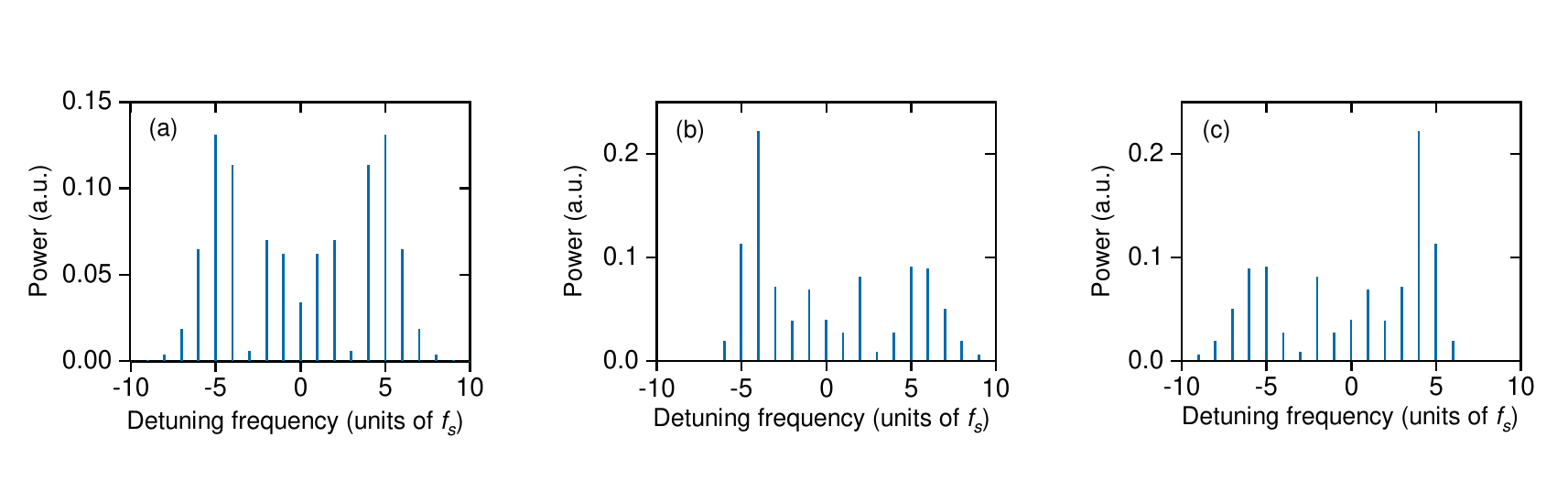}
\caption{\label{fig:2_2_wide}Examples of simulated EO-OFCs. (a) Symmetric spectrum about the carrier frequency when the modulated phase is a single-harmonic signal of frequency $f_s$ with a modulation index $\beta_{1}=2\pi$. (b) Asymmetric spectrum when the modulation phase is a signal composed of two harmonics $f_s$, $2f_s$ with modulation indexes $\beta_{1}=2\pi$, $\beta_{2}=\beta_{1}/10$ and a relative phase of $\varphi_{2}=0$. (c) Spectrum generated using the same modulation parameters as in (b) but with a relative phase $\varphi_{2}=\pi$.}
\end{figure*}

\subsection{\label{sec:level2_2}Reconstruction algorithm}
Consider a linear spectroscopic sample with a spectral response $x(\nu)$, which describes how an input power spectrum is scaled upon interaction with the sample as a function of the optical frequency $\nu$. The measured power $b_{i}$ for the \textit{i}-th probe comb spectrum $A_{i}(\nu)$ can then be expressed as

\begin{equation}
\int^{\nu_{2}}_{\nu_{1}}A_{i}(\nu)x(\nu)d\nu = b_{i}, \\
\label{Eq_2_9}
\end{equation}

\noindent where $\nu_{1}$ and $\nu_{2}$ are, respectively, the limits of the optical bandwidth. After repeating the measurements multiple times, the detection process can be expressed in a discrete form as (see Fig. 1)

\begin{equation}
\textbf{Ax} = \textbf{b}. \\
\label{Eq_2_10}
\end{equation}

In this equation, $\textbf{A}$ is a matrix formed by consecutive probe combs arranged as rows. The element $A_{mn}$ is the spectral power of the \textit{n}-th comb line for the \textit{m}-th input spectrum. The vector $\textbf{x}$ represents the spectral response at the frequency of each comb line, while $\textbf{b}$ denotes the sequence of powers measured during the data acquisition process. The limits $\nu_{1}$ and $\nu_{2}$ correspond to the frequencies of the first and last comb lines, respectively. From now on, $\textbf{A}$ will always be considered a square matrix (that is, $\textbf{x}$ and $\textbf{b}$ will have the same length) to simplify the mathematical treatment.

The most straightforward way to solve the linear system described above is by multiplying the inverse of $\textbf{A}$ by $\textbf{b}$. However, as in many other applications, directly computing the inverse matrix leads to numerical errors or is simply not possible. To clarify this point, it is useful to employ the singular value decomposition (SVD) of $\textbf{A}$, which allows us to express this matrix as $\textbf{A}=\textbf{U}\textbf{S}\textbf{V}^{T}$, where $\textbf{U}$ and $\textbf{V}$ are orthogonal matrices, $T$ denotes the transpose, and $\textbf{S}$ is a diagonal matrix with positive real elements $S_{nn}=s_{n}$ (the singular values of $\textbf{A}$). By convention, the values of $s_{n}$ are arranged in decreasing order. In our approach, the comb spectra generate a system where the ratio between the largest and smallest singular values is extremely high, making direct inversion of $\textbf{A}$ impractical. This issue can be addressed using a "truncated" SVD, where all diagonal elements below a given threshold $s_{th}$ are intentionally set to zero \cite{bibitem35}. This operation facilitates the calculation of the (Moore-Penrose) "pseudo-inverse" matrix, given by $\textbf{A}^{\dagger}=\textbf{V}\textbf{S}^{\dagger}\textbf{U}^{T}$, where $\textbf{S}^{\dagger}$ contains the reciprocals of nonzero elements in the truncated version of $\textbf{S}$. The reconstructed sample response is then calculated as $\textbf{x}=\textbf{A}^{\dagger}\textbf{b}$. It can be proven that the solution $\textbf{x}$ obtained in this way minimizes the Euclidean norm $||\textbf{Ax}-\textbf{b}||_{2}$ \cite{bibitem36}.  

To validate our approach, we conduct a series of numerical simulations based on the previous mathematical analysis (Fig. 3). Although specified in the text, the parameter values used in these simulations and in the experiments of Section III are summarized in Table I at the end of the paper. In a first simulation, we consider a spectral signature similar to the reflectivity of a fiber Bragg grating (FBG). This curve is plotted using an analytical expression derived from coupled-mode theory \cite{bibitem37}. The grating parameters are chosen so that the reflection curve has clearly visible lateral sidelobes, while the width of the central peak is approximately $12$ GHz. Assuming that the filter response $x(\nu)$ is symmetric with respect to the laser frequency, the optical phase can be chosen to be a pure sinusoidal signal with an amplitude modulated according to a staircase function (see Fig. 3a). The length of each step $\Delta t$ is much longer than the inverse of the RF frequency $f_{s}$, $\Delta t \gg 1/f_{s}$. In our simulation, $f_{s}=2$ GHz and $\Delta t=10~\mu$s. The modulation index is changed linearly, so $\beta/\pi$ takes evenly spaced values from $0$ to a maximum level of $4$. Such a high modulation depth can be achieved experimentally by using a low-$V_{\pi}$ PM capable of handling high RF power \cite{bibitem38}. The total number of steps, including an "off-state" measurement in which no signal is sent to the modulator, is $M=31$. After computing the sequence of generated OFCs, we calculate each detected power $b_{i}$ from the discretized version of Eq. (9), normalized to the power of the off-state measurement. Considering that the primary source of noise originates in the detection stage, we simulate a more realistic scenario adding white Gaussian noise to the vector $\textbf{b}$ with a standard deviation $\sigma_{n}=0.003$, similar to the maximum value observed in our experimental results (see Sec. III). 

 To determine the threshold $s_{th}$ of the truncated SVD, we plot the singular values of $\textbf{A}$ as a function of their diagonal index on a logarithmic scale. As expected, they follow a decreasing smooth slope, which becomes significantly more abrupt from a certain element of $\textbf{S}$. Setting all subsequent values equal to zero provides a matrix that minimizes the root mean square error (RMSE) of the reconstruction. Once the threshold parameter is optimized, and as long as the set of programmed combs remains identical, the truncated SVD can be applied to other spectral responses. More details about this procedure can be found in Sec. S1 of the Supplementary Material. The blue squares in Fig. 3(b) are the values calculated from the Moore-Penrose matrix, while the solid line represents the theoretical sample's response. For these results, RMSE $=0.018$ when $s_{th}=0.01$. As can be observed, the reconstructed values accurately reproduce the central peak of the filter response, as well as two sidelobes on the left and right of the central frequency.

The next example is a spectral response resulting from the overlap of two Gaussian absorption curves, one blue-shifted (centered at $6$ GHz) and the other red-shifted (centered at $-2$ GHz) relative to the laser frequency. Each curve has a different height and full width at half maximum (FWHM) (approximately $8$ GHz and $4$ GHz, respectively). In this case, using a driving signal like that in Fig. 3(a) leads to a numerical solution of Eq. (10) that has no physical meaning, since the reconstruction is always symmetric with respect to the laser frequency. This issue can be addressed by applying a two-harmonic signal to the PM to generate asymmetric probe combs (see Fig. 2). As in the previous example, we set the parameter $\beta_{1}/\pi$ (for the frequency $f_{s}$) to take values from 0 to a maximum level of $4$. The values of $\beta_{2}/\pi$ for the $2f_{s}$ signal are chosen to be identical to $\beta_{1}/\pi$ but $10$ times weaker, thus avoiding the practical need to amplify this additional component. We consider $f_{s}=1$ GHz and $M=41$. The specific values of $\beta_{2}$ and the relative phase $\varphi_{2}$ for each PM reconfiguration are selected as follows. We define a vector of values for $\beta_{2}/\pi$, which, as explained above, is formed by a set of $M$ equidistant levels from $0$ to $0.4$. The phase $\varphi_{2}$ forms another vector of the same length as $\beta_{2}/\pi$ constructed by alternating the values $0$ and $\pi$. Next, we randomly permute the elements of both sequences and use the resulting vectors to successively select the parameters of the $2f_{s}$ component at each measurement step. Introducing such a randomness into Eq. (10) facilitates the creation of dissimilar probe spectra \cite{bibitem39}. However, depending on the available electronic equipment, the choice of the driving signal can be simplified, as will be discussed in Sec. III. A sketch of the optical phase generated for three measurement steps is shown in Fig. 3(c). As in the previous example, we add Gaussian noise to the powers $\left\{ b_{i} \right\}(\sigma_{n}=0.003)$ and solve Eq. (10) using a truncated SVD of $\textbf{A} (s_{th}=0.05)$. Fig. 3(d) shows that the recovered $x(\nu)$ fits the reference curve accurately (RMSE $=0.016$). 

\begin{figure}
\centering
\includegraphics[height=6.6cm]{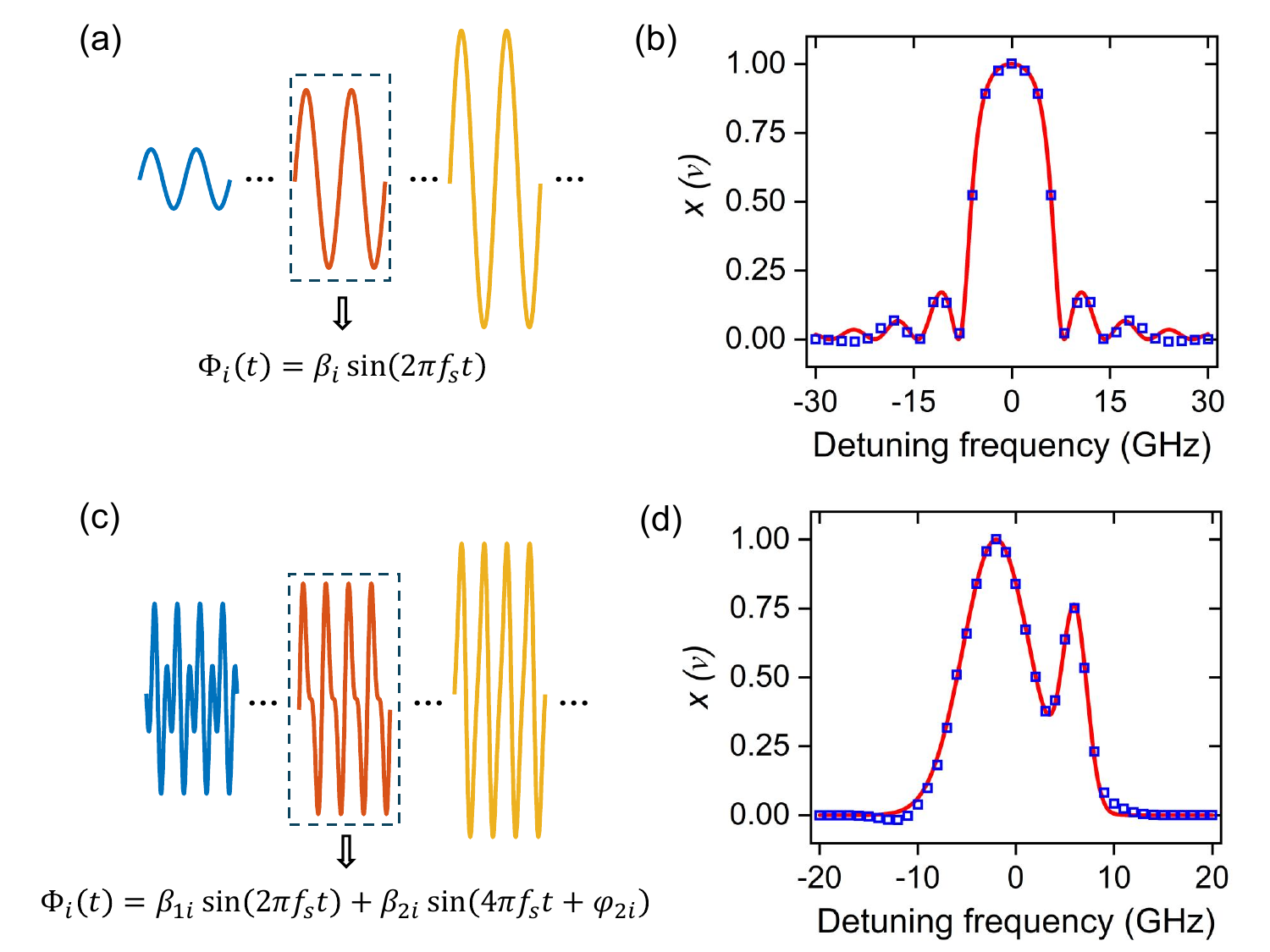}
\caption{\label{fig:3}Numerical simulations. (a) Phase modulation to reconstruct a spectral response similar to the reflectivity of a FBG centered at the laser frequency. For the \textit{i}-th configuration, $\Phi_i(t)$ is a single-tone signal of frequency $f_s$ with a modulation index $\beta_i$. (b) Reconstructed points of the filter response $\textit{x}(\nu)$ (blue squares) from $M=31$ noisy measurements using the method based on the Moore-Penrose matrix. The solid line is the theoretical curve for the FBG reflectivity. (c) Phase modulation employed for the reconstruction of a more complex spectral response, which results from the overlapping of two different Gaussian curves asymmetrically shifted from the laser frequency. For the \textit{i}-th configuration, $\Phi_i(t)$ is a two-tone signal with modulation indices $\beta_{1i}$ and $\beta_{2i}$ corresponding to the frequencies $f_s$ and $2f_s$, respectively. The angle $\varphi_i$ denotes the relative phase between the two harmonics. (d) Reconstructed points of $\textit{x}(\nu)$ (blue squares) and theoretical curve (solid line). In this case, we consider $41$ measurements and, as before, we employ a reconstruction algorithm based on the pseudoinverse matrix method.}
\end{figure}


\section{EXPERIMENTAL RESULTS}
The optical setup used in our experiments is depicted in Fig. 4. As a light source, we employ a continuous-wave laser with a linewidth below $0.1$ kHz, an output power of approximately $20$ mW, and a thermal tuning range of $1$ nm centered around 1545 nm (Koheras Basik E15, NKT). The laser light passes through a phase modulator, PM (MPZ-LN-10, iXblue), working as a reconfigurable comb generator. The modulator has a nominal bandwidth of 10 GHz, which actually extends up to 16 GHz, which corresponds to the maximum attainable comb line spacing. The driving RF signal is produced by a low-cost frequency generator (SynthHD PRO, Windfreak Technologies) and boosted by an RF amplifier (ABP0-0300-01-3830, Wenteq Microwave). Although the damage threshold for the modulator is around 2 W, our RF amplifier limits the maximum applicable RF power to somewhat below 1 W.  Properly selected RF bandpass filters are inserted after the amplifier to suppress harmonics introduced during amplification. The optical signal at the output of the PM is directed to a spectral sample and subsequently detected by a free-space low-frequency amplified photodiode (PDA20CS-EC, Thorlabs), located after a fiber collimator and a focusing lens (not shown in Fig. 4). The detector signal is digitized using an 8-bit oscilloscope (MSO8104, Rigol). Although all the results presented in this section are obtained without optical amplification, if necessary, an amplifier may be added after the comb generator. The laser source and the comb generator are implemented using polarization-maintaining (PM) fiber. This part of the system can be easily integrated into a dual-comb scheme that allows us to obtain a reference curve for each sample.

\begin{figure}
\includegraphics[height=6.0cm]{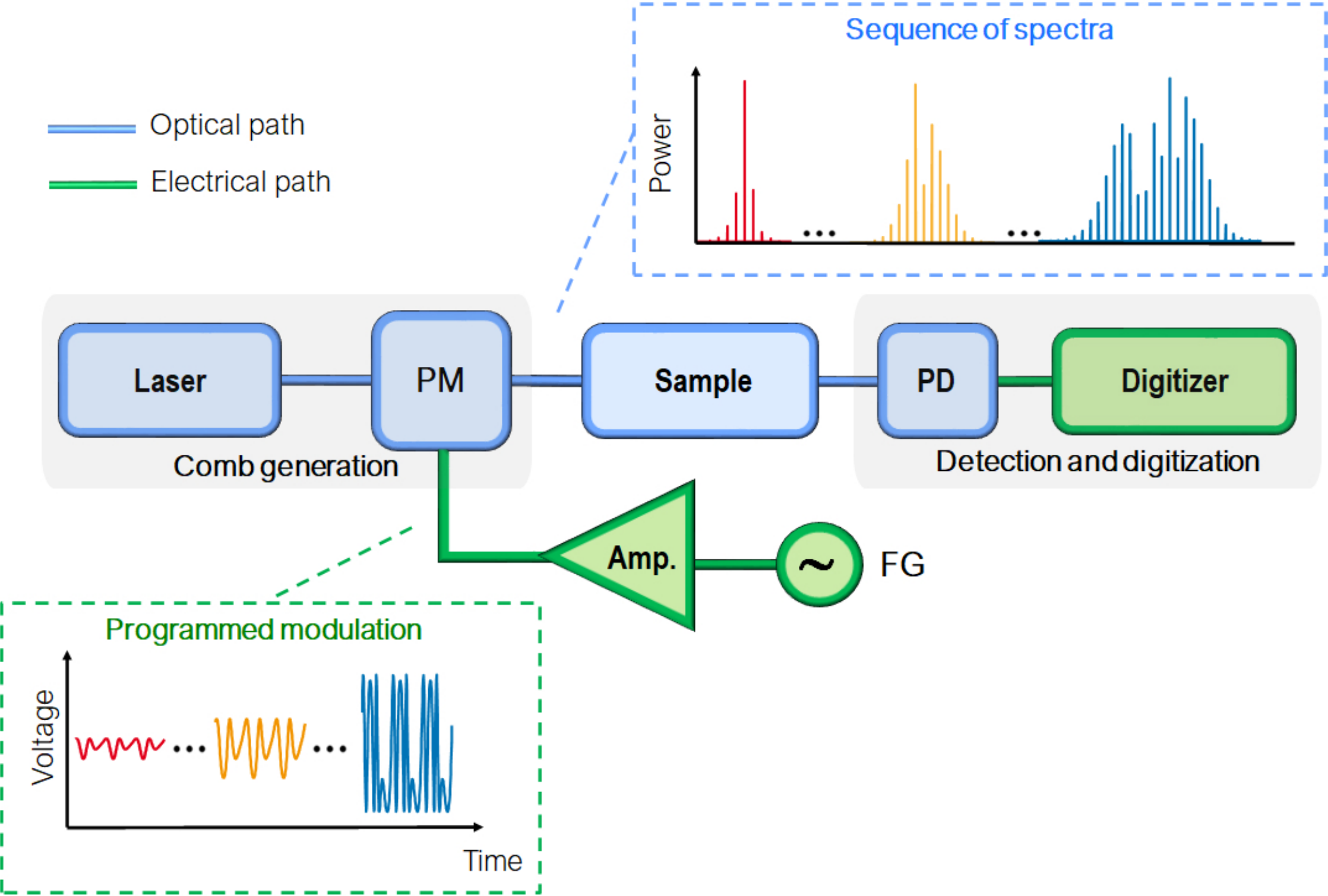}
\caption{\label{fig:4}Sketch of the basic setup for computational EO comb spectroscopy. In the simplest configuration, a phase modulator (PM), fed by a cw laser, is driven by a frequency generator (FG) or an arbitrary waveform generator. The programmed modulation is boosted by an RF amplifier (Amp.). RF filters can be placed at the output of the amplifier to eliminate spurious harmonics (not shown). The sequence of combs generated by the reconfigurable EO modulation interrogates a spectroscopic sample. The light transmitted (or reflected) is captured by a photodetector (PD) and the generated electrical signal is subsequently digitized. To increase the number of comb lines, several PMs can be placed in cascade and the optical signal can be amplified if necessary.}
\end{figure}

As a first example of a spectral sample, we consider a tunable bandpass filter (XTM-50, EXFO). Its transmission curve is adjusted to have a FWHM below the resolution of the optical spectrum analyzer available in our laboratory ($\sim0.05$ nm). The filter response is precisely measured using a dual-comb interferometer (DCI) in a collinear configuration (see Appendix B) with a line spacing of $1$ GHz. The resulting dual-comb points are fitted with a cubic spline curve, serving as a reference. For this measurement, the laser wavelength is aligned with the center of the filter response. The addressing signal is an amplitude-modulated (AM) sinusoid, as shown in Fig. 3(a). The time length $\Delta t$ is $200~ \mu$s, the minimum available when the RF generator is configured in a "sweep" mode. The half-wave voltage at $1$ GHz is determined via a self-heterodyne method \cite{bibitem40}, yielding $V_{\pi}=5.1~V$. By fine-tuning the signal powers programmed in the generator, we linearly change the amplitude of the applied voltage from $1~V$ to $7.6~V$ in 24 steps, which implies varying the value of $\beta / \pi$ from $0.2$ to $1.5$. The acquisition time $t_{m}$ is therefore $4.8$ ms. We add a first measurement for which $\beta=0$. The series of voltages at the modulator RF input are accurately obtained from the measurements taken by a power meter (MA24126A, Anritsu). From the experimental values of $\beta / \pi$, the probe spectra are numerically evaluated. These computed spectra are represented on a linear scale in Sec. S2 of the Supplementary Material. The detected signal is composed of a series of consecutive dc segments, each corresponding to a given modulator configuration. We average $5\times10^{4}$ points per segment, with a standard deviation $\sigma_{n}=0.0015$, nearly constant throughout the oscilloscope trace. The resultant values are normalized using the reference measurement (when no signal is applied to the PM). The elements of the vector $\textbf{b}$ are shown in Fig. 5(a). The reconstructed sample's response $x(\nu)$ is obtained through a truncated SVD ($s_{th}=0.05$) and compared to the reference curve spanning over $16$ GHz (RMSE$=0.018$), as can be observed in Fig. 5(b). 

\begin{figure}
\includegraphics[height=3.0cm]{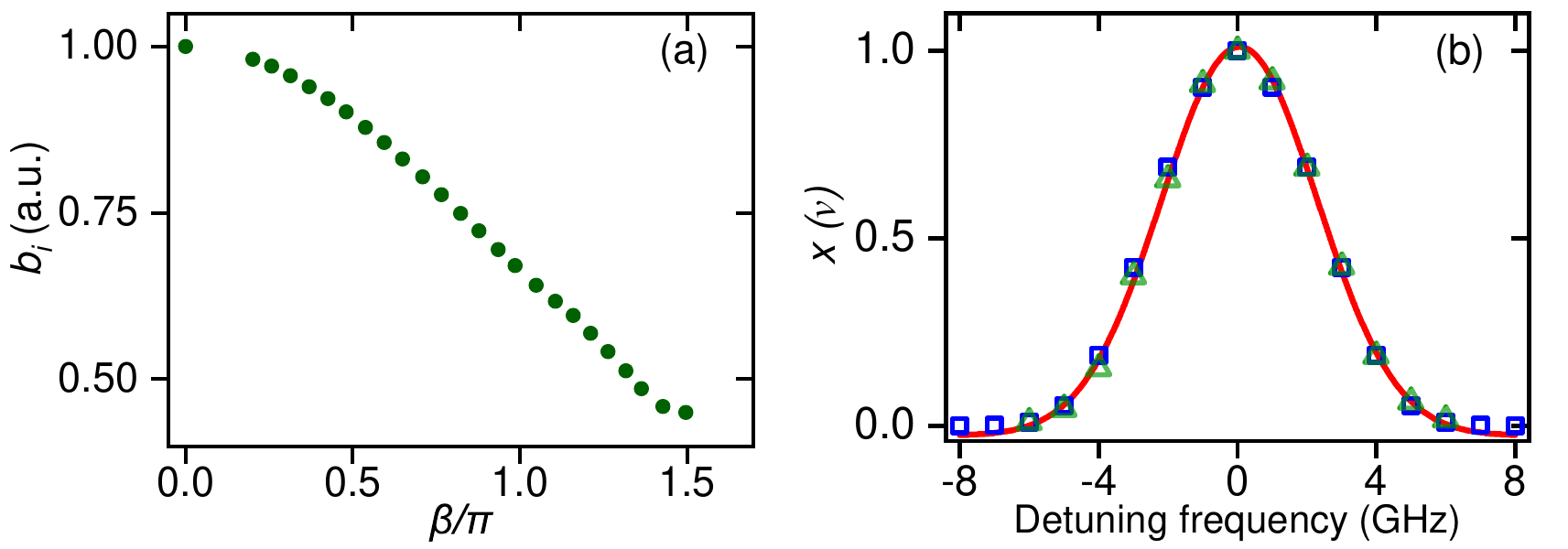}
\caption{\label{fig:5}Experimental results for the transmission $x(\nu)$  of a tunable bandpass filter centered at the laser frequency. (a) Series of measured powers $b_i$ as a function of the modulation index $\beta / \pi$. (b) Reconstructed values of {$x(\nu)$} (blue squares) together with a reference curve (red line) obtained by fitting dual-comb measurements (green triangles) with a Gaussian function. The RMSE of the reconstruction is $0.018$.}
\end{figure}

In the previous result, we have assumed precise alignment between the laser frequency and the center of the filter transmission, which is achievable because of the fine-wavelength tuning of our laser. However, this requirement can be relaxed by employing a two-tone driving signal. With the electronic equipment available in our lab, we can accurately control the relative phase $\varphi_{2}$ between an amplified AM signal ($f_{s}=500$ MHz) and a sinusoid ($2f_{s}=1$ GHz) using an electrical phase shifter operating at $500$ MHz (JSPHS-661+ evaluation board, Mini-Circuits). We consider only two values of $\varphi_{2}~$($0$ and $\pi$), as in the spectra shown in Fig. 2. In this case, the PM driven by the highest RF power provides a maximum of 14 comb lines within 20 dB, requiring us to reduce the filter’s FWHM to the minimum attainable value (around $4$ GHz). Despite all these limitations, this second experiment allows us to reconstruct the filter response when there is a small relative shift between its center and the laser frequency. The voltage amplitude of the $500$ MHz signal is linearly varied (from $0.9$ to $7.6$ V). The half-wave voltage at this frequency is $5$ V. The power of the $1$ GHz sinusoid (coming from the second output of our signal generator) is set to a constant value that does not require amplification ($1$ V of peak amplitude). We consider $51$ reconfigurations, $26$ of them with $\varphi_{2}=0$ (including a measurement with the PM in the off-state) and the remaining $25$ with $\varphi_{2}=\pi$. This driving scheme can be implemented by sending two consecutive modulated signals, each with a different value of $\varphi_{2}$. The numerically computed comb spectra are shown in Sec. S3 of the Supplementary Material. The laser frequency $\nu_{0}$ is shifted $300$ MHz ($0.6f_{s}$). The measured powers $\{ b_{i} \}$ (Fig. 6(a)) are the result of averaging $5\times10^{4}$ points ($\sigma_{n}=0.003$). These measurements demonstrate that the two values of $\varphi_{2}$ (corresponding, respectively, to the dark and light green points in Fig. 6(a)) make it possible to distinguish between the left side and the right side of the filter response with respect to $\nu_{0}$. The reconstructed transmission over $12$ GHz ($s_{th}=0.05$) is shown in Fig. 6(b), along with the reference curve obtained by interpolating the points measured by DCI using a line spacing of $1$ GHz (RMSE $=0.070$). A zoom-in view of the maximum transmission area (inset in Fig. 6(b)) shows again a good fit between both curves, clearly revealing the frequency shift with respect to the laser frequency. Experimentally, one can observe that if $\nu_{0}$ shifts toward the center of the transmission curve, the dark and light green points in Fig. 6 overlap closely, resulting in a single curve, as expected.

\begin{figure}
\includegraphics[height=3.0cm]{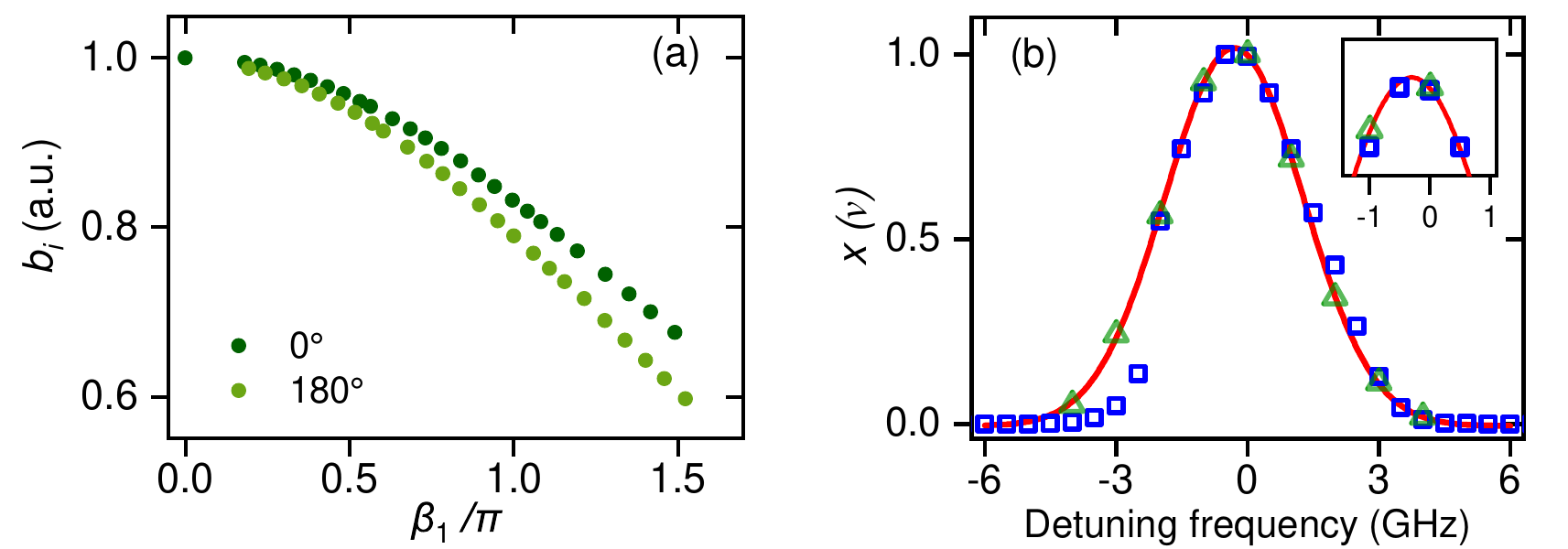}
\caption{\label{fig:6}Experimental results for the transmission $x(\nu)$ of a tunable bandpass filter shifted with respect to the laser frequency. (a) Series of measured powers {$b_{i}$} as a function of the modulation index $\beta_{1}/\pi$ for two values ($0$ and $180°$) of the relative shift between the signals at $f_{s}$ and $2f_{s}$. (b) Reconstructed values of {$x(\nu)$} (blue squares) together with a reference curve (red line) obtained by fitting dual-comb measurements (green triangles) with a Gaussian function. The RMSE of the reconstruction is $0.07$. The inset shows a zoomed-in view of the peak maximum. }
\end{figure}





As a final example, we consider a molecular transition line as a spectroscopic sample. The transmission $T$ for such a fingerprint signature appears as a dip on a constant background at a well-defined frequency. In contrast, the power of the lines of our probe combs eventually tends to zero as they move far enough from the central frequency. Consequently, in this case, the EO combs generated by our system are not suitable sampling functions. This issue can be addressed by removing the background (i.e., the signal in the absence of transition), which leads to the reconstruction of a function with a behavior like that in the previous examples. A similar strategy is employed in compressive dual-comb spectroscopy to ensure sparsity in the spectral responses \cite{bibitem41}. For this purpose, we take an additional power measurement $b_{0}$ far from the center of the molecular transition. The curve $x'=1-T$ is then reconstructed using the linear system of Eq. (10), but with a modified sequence of powers $\{ b'_{i} \}$, where $b'_{i}=1-b_{i}/b_{0}$ (see Appendix C).

In our experiment, we use a gas cell with an absorption length of $5.5$ cm that contains $\text{H}^{\text{13}}\text{CN}$ at $3.3$ kPa ($25$ Torr), with a $10\%$ uncertainty in pressure (Wavelength References). For such a low pressure, a good reconstruction of the absorption line under consideration (P4 at $1545.23$ nm) requires a sampling comb with a line spacing of (at most) $500$ MHz and an optical bandwidth extending on the order of $10$ GHz. In our setup, a single phase modulator cannot meet the above requirements, as the maximum achievable modulation index (approximately $1.7\pi$) is insufficient to generate the required number of comb lines. To overcome this limitation, we implement a comb generator consisting of two cascaded PMs. When driven by in-phase sinusoidal signals of identical frequency, both PMs behave as a single modulator with $\beta=\beta_{1}+\beta_{2}$, where $\beta_{1}$ and $\beta_{2}$ are the modulation indices of each PM \cite{bibitem42}. For this experiment, we align the laser with the transition frequency and use a single-tone AM signal (like in Fig. 3(a)). This driving signal is split into two arms and our phase shifter is inserted in one of them to ensure precise synchronization between both PMs. In each path, identical RF amplifiers are inserted (ABP0-0300-01-3830, Wenteq Microwave). The half-wave voltage at $500$ MHz for the second PM is $4.9$ V. The values of $\beta_{1}/\pi$ and $\beta_{2}/\pi$ undergo an approximately linear variation from $0.18$ to $1.64$ and from $0.15$ to $1.68$, respectively. With a sufficiently high modulation index, a single PM could be employed to generate the combs, while the phase shifter would assist in producing a two-tone modulation, thereby eliminating the need to align the laser with the center of the absorption line (see Sec. II). As in the previous experiment, we consider $51$ reconfigurations (including an off-state measurement), so the acquisition time is $10$ ms, see Fig. 7(a). Each $b'_{i}$ is the result of averaging $5\times10^4$ points ($\sigma_{n}=0.001$). By solving the corresponding linear system ($s_{th}=0.03$), we reconstruct $x'(v)$. As a reference, we used the P4 line shape for the HCN at $25$ Torr, derived from the NIST data. The reconstructed points spanning over $13$ GHz, shown in Fig. 7(b), closely match the reference values (RMSE $=0.009$).

\begin{figure}
\includegraphics[height=3.0cm]{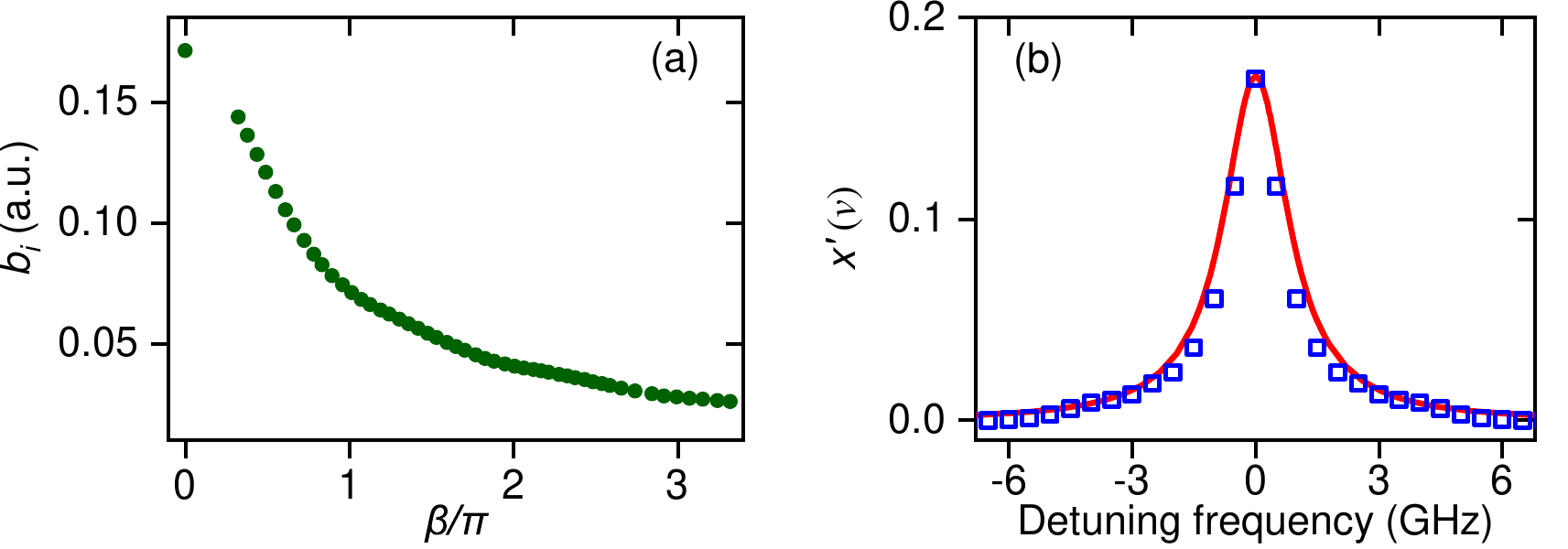}
\caption{\label{fig:7}Experimental results for the response $x'=1-T$, where T is the transmission of a $5.5$ cm-long HCN gas cell at $25$ Torr, centered at $1545.23$ nm. (a) Series of the measured powers $b_i$ as a function of the modulation index $\beta / \pi$. (b) Reconstructed spectral values of $x'$, (blue squares) and the reference curve (red line) obtained from NIST data. The RMSE of the reconstruction is $0.009$.}
\end{figure}

	

\renewcommand{\arraystretch}{2.4} 
\begin{table*}[t]
\caption{Summary of parameter values from simulations and experiments}	
    \centering
    \begin{tabular}{@{} lccccccc @{}}
        \toprule
        \multicolumn{1}{c}{} & 
        \textbf{\boldmath$f_s$ (GHz)} & 
        \textbf{\boldmath$V_\pi$ (V)} & 
        \textbf{\boldmath$\beta_{1,\text{max}}/\pi$} & 
        \textbf{\boldmath$\beta_{2,\text{max}}/\pi$} & 
        \textbf{M} & 
        \textbf{\boldmath$t_m$ (ms)} & 
        \textbf{RMSE} \\
        \midrule
        \textbf{Fig. 3(b)} & 2   & 3   & 4    & --  & 31 & 0.3  & 0.018 \\
        \textbf{Fig. 3(d)} & 1   & 3   & 4    & 0.4 & 41 & 0.4  & 0.016 \\
        \textbf{Fig. 5}  & 1   & 5.1 & 1.50 & --  & 25 & 4.8  & 0.018 \\
        \textbf{Fig. 6}  & 0.5 & 5    & 1.52 & 0.2 & 51 & 10   & 0.070  \\
        \textbf{Fig. 7}  & 0.5 & \makecell{5 (PM1)\\4.9 (PM2)} 
                        & \makecell{1.64 (PM1)\\1.68 (PM2)} 
                        & -- 
                        & 51 
                        & 10 
                        & 0.009 \\
        \bottomrule
    \end{tabular}
 
\end{table*}

\section{\label{sec:level4}DISCUSSION AND CONCLUSIONS}
In this paper, we have presented a computational method for spectroscopic measurements based on a rapidly reconfigurable EO comb generator fed by a narrow-linewidth cw laser. The capability of this system to deal with different classes of spectroscopic samples has been demonstrated by numerical simulations. Despite limitations imposed by the available electronic equipment, the experimental results obtained in our laboratory confirm the feasibility of the proposed approach. We have successfully reconstructed the transmission of an optical bandpass filter with a FWHM below the resolution of a typical grating-based optical spectrum analyzer ($<0.05$ nm). We have also reconstructed a molecular absorption line of HCN with a frequency sampling of $500$ MHz covering $>10$ GHz. For this particular measurement, we have used two phase modulators in cascade (to increase the number of attainable comb lines). However, this two-stage configuration could be avoided with an ultra-low-$V_\pi$ phase modulator capable of handling high RF power \cite{bibitem38}.The number of system reconfigurations was chosen to keep the acquisition time within 10 ms, although this number could be reduced to speed up acquisition, as long as the generated spectra cover a sufficiently broad optical bandwidth and the system to be solved remains square (see Section S4 of the Supplementary Material). Additionally, the measurement speed could be further increased with faster reconfigurable electronics. 

Compared with traditional dual-comb or interferometric techniques, our method reduces spectroscopic system complexity while preserving the high spectral sampling of frequency combs. In our experiments, we employed a highly stable ultra-narrow-linewidth laser ($<0.1$ kHz). Nevertheless, given the line spacings considered (hundreds of MHz to a few GHz), more economical sources with broader linewidths, such as distributed feedback (DFB) lasers, could still yield comb spectra reasonably close to the ideal noise-free case. Although optical signal amplification was not required in this work, in practical scenarios with significant losses it may become necessary. In such cases, the perturbation introduced by amplified spontaneous emission noise in the sampling spectra would need to be evaluated, and the truncated inverse technique might need to be replaced or complemented by more advanced reconstruction algorithms \cite{bibitem43}.  

We have considered signals with up to two harmonics, while Appendix A outlines the general framework for multi-tone modulation ($h>$2). Despite being mathematically more involved, this approach increases the degrees of freedom to control the shape of the comb spectra, a feature already exploited to flatten EO combs, both resonant \cite{bibitem44} and non-resonant \cite{bibitem9}. Furthermore, by programming the modulation index and phase of each harmonic, the optical power of individual comb lines can be tailored through optimization algorithms, an approach recently applied to integrated parallel-convolution processing \cite{bibitem45}. This opens up the possibility of addressing complex spectroscopic responses and optimizing the reconstruction algorithm, albeit at the cost of more sophisticated driving signals, generated, for example, by ultra-high-speed AWGs operating in the GHz range.

We have focused on phase modulation and showed that a wide range of comb spectra can be produced using multitone RF signals. The potential extension to more general EO comb configurations—including both phase and intensity modulators—is theoretically straightforward. Such multi-stage generators could increase several times the number of comb lines (and, hence, the bandwidth) beyond the maximum value demonstrated here ($M\leq27$ within 20 dB of power). Additionally, a Mach-Zehnder modulator (MZM) driven by an AWG could be used to map custom-designed digitally generated spectra into the optical domain \cite{bibitem31}. These extensions to other configurations—or even a change in the comb platform—are possible provided that two basic requirements are met: the ability to dynamically reconfigure the comb generator in a predictable or measurable manner, and the generation of sufficiently dissimilar and stable combs. Experimentally, the sampling spectra may be measured directly—for a fixed driving configuration—using heterodyne interferometry \cite{bibitem46} or a ultra-high-resolution spectrum analyzer.

From a mathematical perspective, the inverse problem solved by our method is similar to that encountered in the design of miniaturized reconstructive spectrometers \cite{bibitem47}. As with such systems, a variety of reconstruction strategies could be employed, ranging from the truncated singular value decomposition shown here \cite{bibitem35,bibitem43} to various optimization techniques that address the ill-posed nature of the linear system \cite{bibitem48,bibitem49}. The optimal reconstruction method depends on both the complexity of the driving signals and the assessment of the different error sources present in our experimental system. This paves the way for further investigation into which strategy yields the most accurate and robust results depending on the targeted application.

Our computational procedure has been demonstrated using a notably simple and easy-to-implement system, which makes it an excellent candidate for integration into a low-complexity photonic circuit consisting of a small number of optical elements \cite{bibitem50}. Integrated lithium niobate (LN) phase modulators, for instance, have rapidly advanced to offer low on-chip losses, low half-wave voltages over broad frequency ranges, and high RF power-handling capabilities (up to several watts), thus enabling modulation indexes as high as $\beta/\pi=4$ \cite{bibitem51}, which could outperform the results shown here. In parallel, tailored RF waveforms could be generated using direct digital synthesizers (DDS) \cite{bibitem52} or field-programmable gate arrays (FPGAs) combined with fast digital-to-analog converters (DACs), offering versatility while substantially reducing overall cost compared to high-end AWGs. The presented approach, therefore, might contribute to the development of integrable and reproducible optical architectures for applications such as remote sensing, gas detection, or hyperspectral imaging, benefiting from the high-frequency resolution of optical frequency combs while maintaining remarkable hardware simplicity.

\section*{Acknowledgments}
This publication is part of the project CNS2023-144732 funded by MICIU/AEI/10.13039/501100011033 and European Union NextGenerationEU/PRTR.  It is also funded by MCIN/AEI/10.13039/501100011033/ “FEDER A way of making Europe” (PID2021-124814NB-C22), Generalitat Valenciana (CIPROM/23/44) and Universitat Jaume I (UJI-B2022-53).

 V. Durán acknowledges support from the grant RYC-2017–23668, funded by MCIN/AEI/10.13039/501100011033, during the initial stage of this work, and thanks P. Clemente and F. Soldevila for their valuable comments and suggestions on the original idea of the paper.

\section*{Author Declaration}
\noindent Conflict of Interest. The authors have no conflicts to disclose.

\section*{Data Availability Statement}
The data that support the findings of this study are available from the corresponding author upon reasonable request.

\appendices

\section{Multi-tone driving signal}

\subsection*{$h$-tone modulation}
In the most general case, the driving signal comprises $h$ harmonics, so that the instantaneous phase of the field $\Phi(t)$ for an arbitrary reconfiguration of our system is given by:

\begin{equation}
\Phi\left( t \right) = \sum^{h}_{l=1}\beta_{l}sin\left( 2\pi l f_{s}t + \varphi_{l} \right), \\
\label{Eq_A_1}
\end{equation}

\noindent which corresponds to equation (3), where $f_{s}$ is the fundamental RF frequency and $\beta_{l}$ and $\varphi_{l}$ are, respectively, the modulation index and the phase of the \textit{l}-harmonic. We have omitted the subscript~$i$ to simplify the notation. Accordingly, the field $E$ can be expressed as:

\begin{equation}
E \left( t \right) = E_{0}e^{j2\pi \nu_{0}t}e^{j\Phi\left( t \right)}, \\
\label{Eq_A_2}
\end{equation}

\noindent where $E_{0}$ is the field amplitude, $\nu_{0}$ is the carrier frequency and $j$ denotes the imaginary unit. This equation corresponds to Eq. (2) in Sec. (II). 

Considering the Jacobi-Anger expansion, the above equation takes the form:

\begin{equation}
\begin{split}
\label{Eq_A_3}
E (t) = E_{0}&e^{j2\pi \nu_{0}t} \\
\prod_{l=1}^{h} \sum_{k_{l}=-\infty}^{\infty}J_{k_{l}}(\beta_{l}) & e^{j2\pi k_{l}lf_{s}t} e^{jk_{l}\varphi_{l}}. \\
\end{split}
\end{equation}

By taking the Fourier transform and slightly expanding the resulting expression, $\tilde{E}(\nu)$ can be written as:


\begin{equation}
\label{Eq_A_4}
\begin{aligned}
\tilde{E}(\nu) =\; & E_{0} \sum_{k_{1}=-\infty}^{\infty} \sum_{k_{2}=-\infty}^{\infty} \dotsb \sum_{k_{h}=-\infty}^{\infty} e^{\, j \left( k_{1}\varphi_{1} + k_{2}\varphi_{2} + \dotsb + k_{h}\varphi_{h} \right)}
 \\[6pt]
&\cdot J_{k_{1}}(\beta_{1}) J_{k_{2}}(\beta_{2}) \dotsb J_{k_{h}}(\beta_{h}) \\[6pt]
& \cdot \delta \!\left( \nu - \nu_{0} - \left( k_{1}+2k_{2}+\dotsb+hk_{h} \right) f_{s} \right). \\[6pt]
\end{aligned}
\end{equation}

Despite its complexity, this equation reflects a fundamental fact: each line in the resulting comb is intricately determined by the amplitudes and phases of the driving harmonics, providing many degrees of freedom to set its value. 
\subsection*{Two-tone modulation}
For the particular case of two tones and according to the theory presented in Ref. [34], it is possible to write a generalization of the Jacobi-Anger expansion in the following form:

\begin{equation}
e^{j\left( xsin\theta + ysin\left(2\theta+\phi \right) \right)} = \sum_{k=-\infty}^{\infty} e^{jk\theta} J_{k}(x,y;e^{j\phi}) , \\
\label{Eq_A_5}
\end{equation}

\noindent where the variables $x,y \in \mathbb{R}$, the angles $\theta,\phi \in [0,2\pi]$ and $J_{k}(x,y;e^{j\phi})$ is a two-variable, one-parameter generalized Bessel function (GBF), defined in terms of ordinary one-variable Bessel functions as:

\begin{equation}
J_{k}(x,y;e^{j\phi}) = \sum_{l=-\infty}^{\infty} \left( e^{j\phi} \right)^{l} J_{k-2l}(x)J_{l}(y) . \\
\label{Eq_A_6}
\end{equation}

The above two expressions can be employed for the \textit{i}-th reconfiguration of a phase modulator driven by a two-tone RF signal through the next substitutions:

\begin{equation}
\theta = 2\pi f_{s}t, \phi = \varphi_{2i}, x= \beta_{1i}, y = \beta_{2i} , \\
\label{Eq_A_8}
\end{equation}

\noindent where $t$ is the time, $f_{s}$ is the driving RF frequency, $\beta_{1i}$ and $\beta_{2i}$ are the modulation indices for each tone and $\varphi_{2i}$ is the relative phase between them. The result of these substitutions are the Eqs.(\ref{Eq_2_7}) and (\ref{Eq_2_8}) of the main manuscript.

The relationship between GBFs of order
$k$ and $-k$ is given by \cite{bibitem34}:

\begin{equation}
\begin{split}
\label{Eq_A_9}
J_{-k}(x,y;e^{j\phi}) = &(-1)^{k}J_{k}(x,y;-e^{-j\phi}) \\ 
= J_{k}(-x&,-y;e^{-j\phi}). \\
\end{split}
\end{equation}



This symmetry property, combined with Eq. (\ref{Eq_A_6}), helps explain the asymmetry with respect to the central frequency observed in the spectra generated by a two-tone signal, as shown in Fig. 2. 

\section{Experimental setup for dual-comb spectroscopy}
The experimental setup used for dual-comb spectroscopy is shown in Fig. 8. The output power ($20$ mV) from a narrow-linewidth laser (Koheras Basik E15, NKT) is split into two arms, each containing an identical phase modulator (MPZ-LN-10, iXblue) driven by an amplified RF signal coming from a frequency generator. The two driving RF signals have slightly different frequencies $(f_{s}=1$ GHz and $f_{s}+\delta f_{s}=1.0001$ GHz). An acousto-optic frequency shifter (AA Opto-Electronic), placed in one arm, shifts the laser frequency by $f_{AO}=80$ MHz to prevent overlap in the down-converted RF spectrum. The two generated optical combs are recombined in fiber and then split again into two paths. In one of them, the light passes through a spectroscopic sample before detection, while the other path contains only a photodetector and serves as a reference. This scheme—known as symmetric or collinear, as both combs pass through the sample—enables the measurement of the sample absorption. Although phase information is not directly accessible in this configuration (it requires additional processing and an extended RF bandwidth \cite{bibitem53}), the optical setup remains robust to external perturbations. Balanced photodetectors are used at the end of both paths, BPD1 (PDB470C, Thorlabs) and BPD2 (PD100B, Koheron), and the detection signals are digitalized by a single oscilloscope (MSO8104, Rigol). We acquire $20$ ms traces, each containing $2000$ consecutive interferograms. The Fourier transform of each trace yields an RF comb with a line spacing of $100$ kHz. The powers applied to the phase modulators are adjusted to minimize amplitude variations in the dual-comb RF spectrum, at the cost of reducing the total number of resolved comb lines to a maximum of $13$. The sample response, encoded in the RF comb measured in the sample arm, is retrieved by normalizing to the spectrum obtained in the reference arm. With the sole exception of the sample itself, all components in the dual-comb setup use polarization-maintaining fiber. Three RF generators are used in the setup: two for driving the phase modulators (DSG821, Rigol; SSG3032X, Siglent), and one for the AOFS (SDG6032X, Siglent). All generators, along with the oscilloscope, are locked to a highly stable reference clock (OCXO B08, Rigol).

\begin{figure}[H]
\centering
\includegraphics[height=3.0cm]{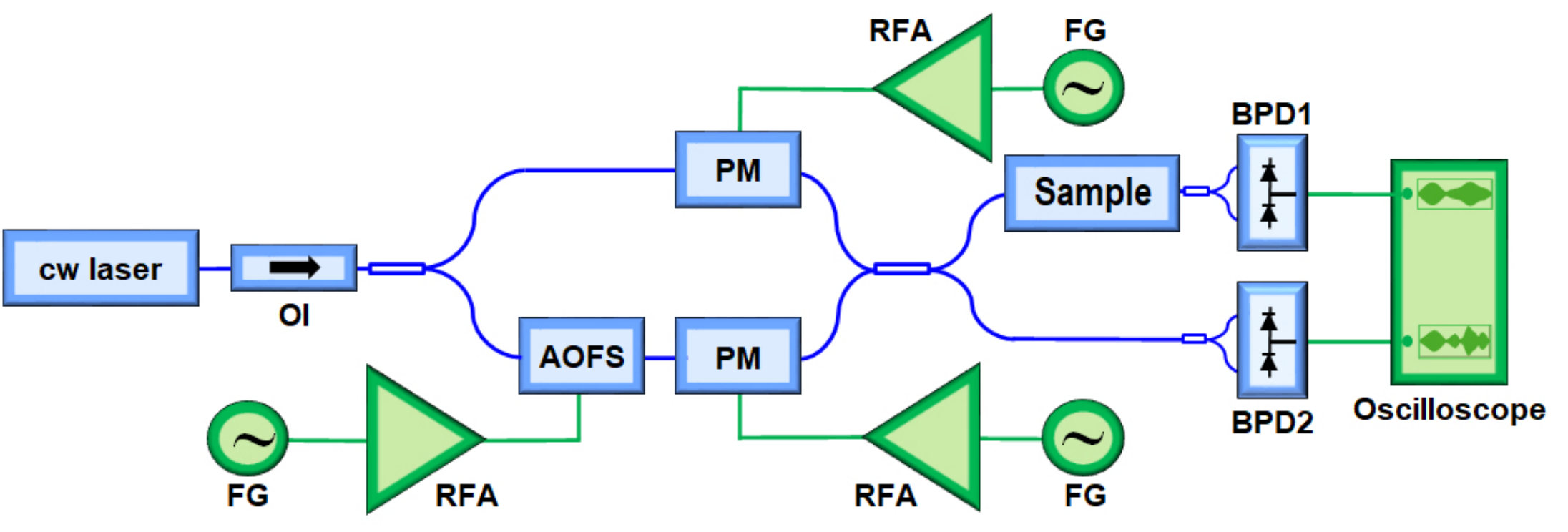}
\caption{\label{fig:B_1}Setup for dual-comb spectroscopy. OI: optical isolator; AOFS: acousto-optic frequency shifter; PM: phase modulator; BPD: balanced photodetector; FG: frequency generator; RFA: radiofrequency amplifier.}
\end{figure}

\section{Inverse problem for the molecular absorption line}
Let $x$ denote our spectroscopic sample, which is the discretized version of a molecular transition line as shown in Fig. 9(a) (typically described by a Voigt profile). Over the considered optical bandwidth, we will assume that the maximum value of the transmission $T(\nu)$ as a function of the optical frequency $\nu$ is $1$. In this case, due to the characteristics of our probe comb, we reformulate the inverse problem to reconstruct $x'=1-T$, instead of $T$, see Fig. 9(b). After discretization, we obtain $\mathbf{Ax'=b'}$, where $\mathbf{A}$ is the square matrix $M \times M$ formed by the consecutive probe combs arranged in rows and the vectors $\mathbf{x'}$ and $\mathbf{b'}$ are, respectively, the sample response and the set of measured power values. According to the definition of ${x'}$, we can write $\mathbf{b'}$ as $\mathbf{b'=A(1_{M}-x)}$, where $1_{M}$ is a column vector consisting of $M$ ones, and $\mathbf{x}$ is the transmission $T$ at each comb line. Assuming that the total power of all comb lines sums to $1$, and using $\mathbf{Ax=b}$ (see Eq.(\ref{Eq_2_10})), the \textit{i}-th component of $\mathbf{b'}$ is simply given by $b'_{i}=1-b_{i}$. If, instead, the plateau transmission is $b_{0}$, the normalized value of $b'_{i}$ is given by $b'_{i}=1-b_{i}/b_{0}$. The constant $b_{0}$ can be determined in several ways, as discussed in the main text and in Ref. \cite{bibitem41}.

\begin{figure}[H]
\centering
\includegraphics[height=3.0cm]{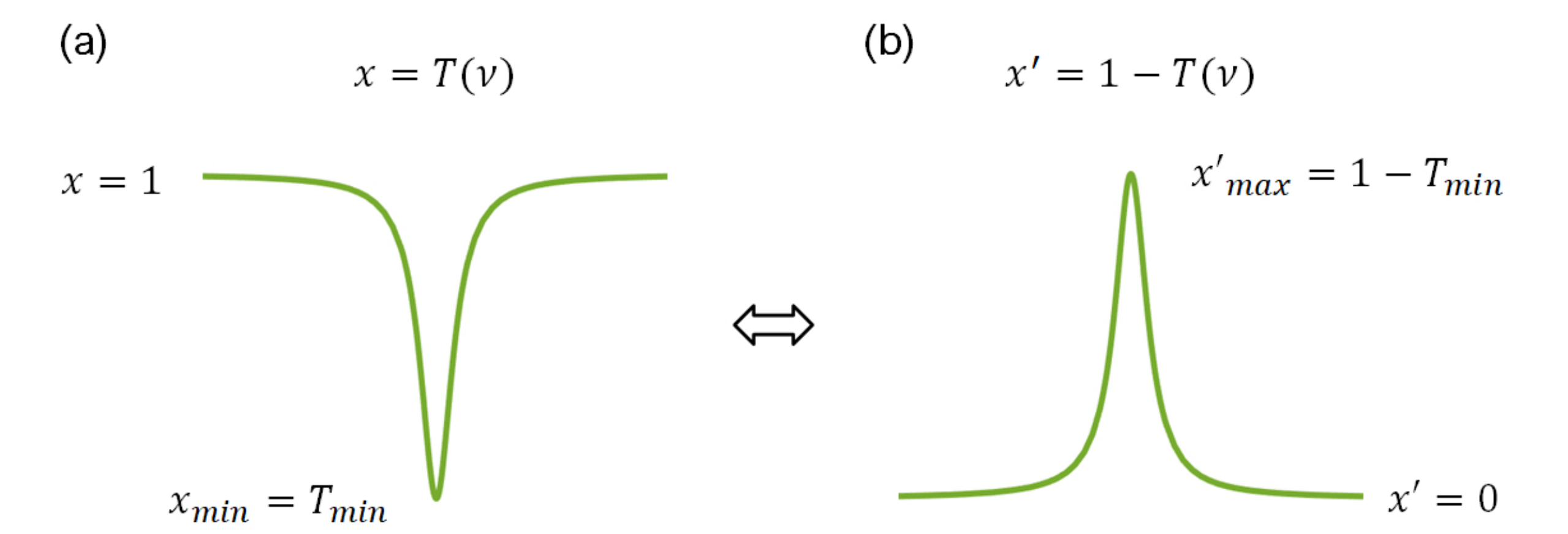}
\caption{\label{fig:C_1} Inverse problem for reconstructing the absorption line. (a) Transmission curve $T(\nu)$ for a molecular transition. It typically follows a Voigt profile, reaching a minimum value $T_{min}$. (b) In our computational approach, we reformulate the inverse problem to retrieve $1-T(\nu)$.}
\end{figure}



\AtBeginEnvironment{thebibliography}{\color{black}}

\vfill

\end{document}